\documentclass[12pt]{article}
\usepackage{graphicx,textcomp}
\usepackage[super,sort&compress,square]{natbib}
\setcitestyle{citesep={,}}

\usepackage[labelfont=bf,justification=raggedright,singlelinecheck=false]{caption}
\usepackage{titlesec}
\usepackage[parfill]{parskip}
\titlespacing{\subsubsection}{0em}{1em}{-1em}

\usepackage{setspace}
\usepackage{geometry}
\begin{document}
\raggedright

\textbf{Electron-Beam Manipulation of Silicon Impurities in Single-Walled Carbon Nanotubes}

\textit{Kimmo Mustonen*, Alexander Markevich, Mukesh Tripathi, Heena Inani, Er-Xiong Ding, Aqeel Hussain, Clemens Mangler, Esko I. Kauppinen, Jani Kotakoski, and Toma Susi*}

Dr. K. Mustonen 1, M. Tripathi 1, Dr. A. Markevich 1, H. Inani 1, Dr. C. Mangler 1, Prof. J. Kotakoski 1, Dr. T. Susi 1\\
1 University of Vienna, Faculty of Physics, Boltzmanngasse 5, 1090 Vienna, Austria\\
E-mail: kimmo.mustonen@univie.ac.at, toma.susi@univie.ac.at

A. Hussain 2, E. Ding 2, Prof. Esko I. Kauppinen 2\\
2 Aalto University School of Science, Department of Applied Physics, P.O. Box 15100, FI-00076 Aalto, Finland


The recent discovery that impurity atoms in crystals can be manipulated with focused electron irradiation has opened novel perspectives for top-down atomic engineering. These achievements have been enabled by advances in electron optics and microscope stability, but also in the preparation of suitable materials with impurity elements incorporated via ion and electron-beam irradiation or chemical means. Here it is shown that silicon heteroatoms introduced via plasma irradiation into the lattice of single-walled carbon nanotubes (SWCNTs) can be manipulated using a focused 55--60 keV electron probe aimed at neighboring carbon sites. Moving the silicon atom mainly along the longitudinal axis of large 2.7~nm diameter tubes, more than 90 controlled lattice jumps were recorded and the relevant displacement cross sections estimated. Molecular dynamics simulations show that even in 2~nm SWCNTs the threshold energies for out-of-plane dynamics are different than in graphene, and depend on the orientation of the silicon-carbon bond with respect to the electron beam as well as the local bonding of the displaced carbon atom and its neighbors. Atomic-level engineering of SWCNTs where the electron wave functions are more strictly confined than in two-dimensional materials may enable the fabrication of tunable electronic resonators and other devices.

\clearpage

Graphene, ideally an infinite monoatomic layer of hexagonally bonded carbon atoms, is a zero bandgap semiconductor in which electrons propagate as massless Dirac fermions.~\cite{novoselov_two-dimensional_2005} Single-walled carbon nanotubes (SWCNTs)~\cite{s._iijima_single-shell_1993} can be envisioned as a cylidrically wrapped section of graphene with connected perimeters. This structural difference confines the electron wave functions on the circumference of the tube, creating a (quasi-)one-dimensional (1D) quantum channel with a band structure dependent on the graphene cutting direction and tube diameter.~\cite{saito_electronic_1992} This 1D nature renders electronic transport in SWCNTs highly sensitive to any perturbations within (or outside) the structure,~\cite{charlier_electronic_2007} which is useful e.g. for chemical sensors. Their electronic transport properties can further be modified by introducing heteroatoms into the graphitic lattice.~\cite{stephan_doping_1994,ayala_physical_2010} Although typically not purposefully introduced, silicon (Si) atoms are often found in as impurities in graphene.~\cite{zhou_direct_2012,ramasse_probing_2013} A high density of Si in both SWCNTs and graphene was recently introduced by simultaneously applying low-energy plasma and laser irradiation, and verified by atomic resolution scanning transmission electron microscopy (STEM).~\cite{inani_silicon_2019}

The effects of electron irradiation on SWCNTs have thus far been considered in the context of knock-on damage. In contrast to graphene, both the chirality and the diameter of the SWCNT has been found to influence displacement threshold energies, with smaller and more reactive tubes being easier to damage.~\cite{krasheninnikov_stability_2005} Further, the curved geometry and the orientation of each atomic site on the tube wall with respect to the electron beam direction makes a full description of the scattering process significantly more complicated.~\cite{zobelli_electron_2007} Finally, tight-binding methods have been used to describe knock-on damage in pristine nanotubes,~\cite{krasheninnikov_stability_2005} but they fail to provide even a qualitatively correct picture for systems with impurities.~\cite{loponen_nitrogen-doped_2006,susi_atomistic_2012} Besides causing damage, focused electron irradiation has recently been recognized as a tool for atom manipulation,~\cite{kalinin_fire_2016,susi_manipulating_2017,susi_towards_2017} complementary to the established capabilities of scanning probe microscopy.~\cite{crommie_confinement_1993,custance_atomic_2009} Until now such manipulation has not been demonstrated in nanotubes.

Here, we show the possibility of moving Si impurities in SWCNTs via an out-of-plane ``bond inversion'' (direct exchange) process,~\cite{susi_siliconcarbon_2014} similar to what was recently achieved in graphene.~\cite{susi_manipulating_2017,dyck_placing_2017,tripathi_electron-beam_2018,dyck_building_2018,su_competing_2018} We show that we are indeed able to manipulate them mainly along the axis of larger diameter tubes, where the atomic structure can be visualized and the local geometry resembles graphene. Using density functional theory based molecular dynamics (DFT/MD) simulations, we further explore the energetics of various dynamic processes as a function of tube diameter and chirality.

For these experiments, the raw material was synthesized~\cite{ding_highly_2017,ding_high-performance_2018} with a floating catalyst process yielding primarily SWCNTs with high chiral angles and a large mean diameter close to 2~nm (see Methods). The Si impurities were incorporated through the use of simultaneous plasma and laser irradiation as described in our recent work,~\cite{inani_silicon_2019} resulting in $\sim$63\% three-fold atomic coordination (Si-C$_3$), the rest being four-fold (Si-C$_4$).~\cite{zhou_direct_2012,ramasse_probing_2013,nieman_structure_2017} To accomplish controlled electron-beam manipulation, the samples were carefully examined to identify near-armchair tubes (due to the sample chirality distribution and to allow atomic resolution imaging) with Si impurities. When a suitable C-Si$_3$ site was identified, the beam was positioned for a fixed period of time over a chosen carbon neighbour, and a frame was acquired after each spot irradiation, as in our recent work with graphene.~\cite{susi_manipulating_2017,tripathi_electron-beam_2018}

\begin{figure}[ht!]
\centering
\includegraphics[width=\textwidth,keepaspectratio]{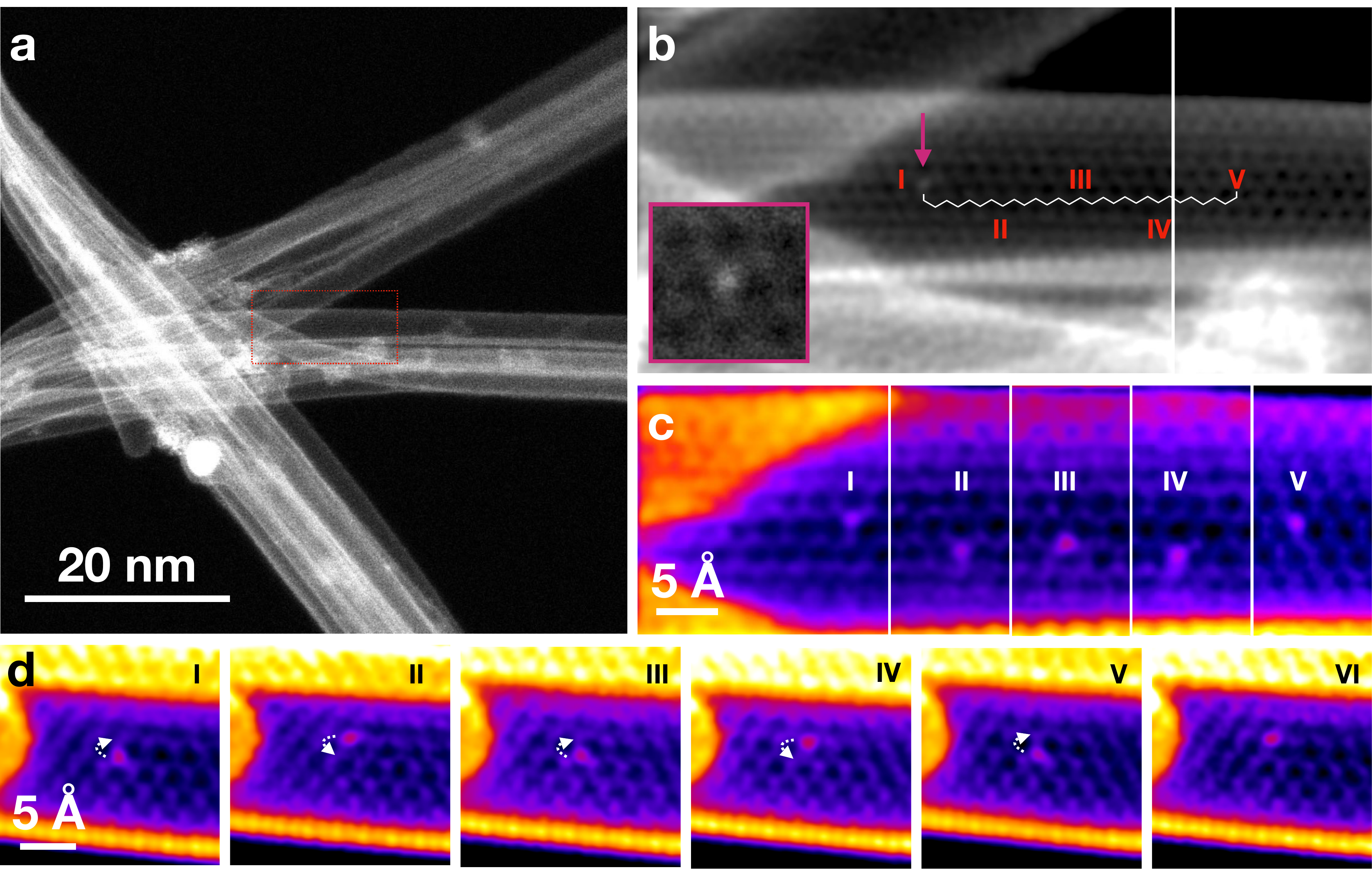}
\caption{Electron-beam manipulation of Si impurities in SWCNTs. (a-c) An atom walked along the zigzag direction of a (20,20) armchair nanotube using focused 60 keV electron irradiation. The intermediate positions are marked with I-V. (d) A controlled back-and-forth movement of an atom in a (22,18) nanotube at 55 keV.}
\label{fig:manipulation}
\end{figure}

The experiments were performed using two primary beam energies, 55 and 60~keV; at 55~keV, a spot irradiation time of 10~s was used whereas at 60~keV the time was reduced to 7~s. \textbf{Figure}~\ref{fig:manipulation}a-c summarizes one Si atom manipulation sequence in a (20,20) armchair tube with a diameter of $\sim$2.75 nm. The atom was directed on a path along the tube axis as shown in Figure~\ref{fig:manipulation}b, covering a total of 30 lattice sites following mainly the zigzag direction. Figure~\ref{fig:manipulation}c shows snapshots of positions I-V separately highlighted in Figure~\ref{fig:manipulation}b. Another experiment consisted of a series of repeated back-and-forth jumps along the armchair direction perpendicular to the axis of a large-diameter near-armchair tube. Figure~\ref{fig:manipulation}d shows the path of the atom on what is likely a (22,18) SWCNT with a diameter of $\sim$2.67~nm, moving repeatedly between the two sublattices in a fully controlled manner.

\begin{figure}[ht!]
\centering
\includegraphics[width=\textwidth,keepaspectratio]{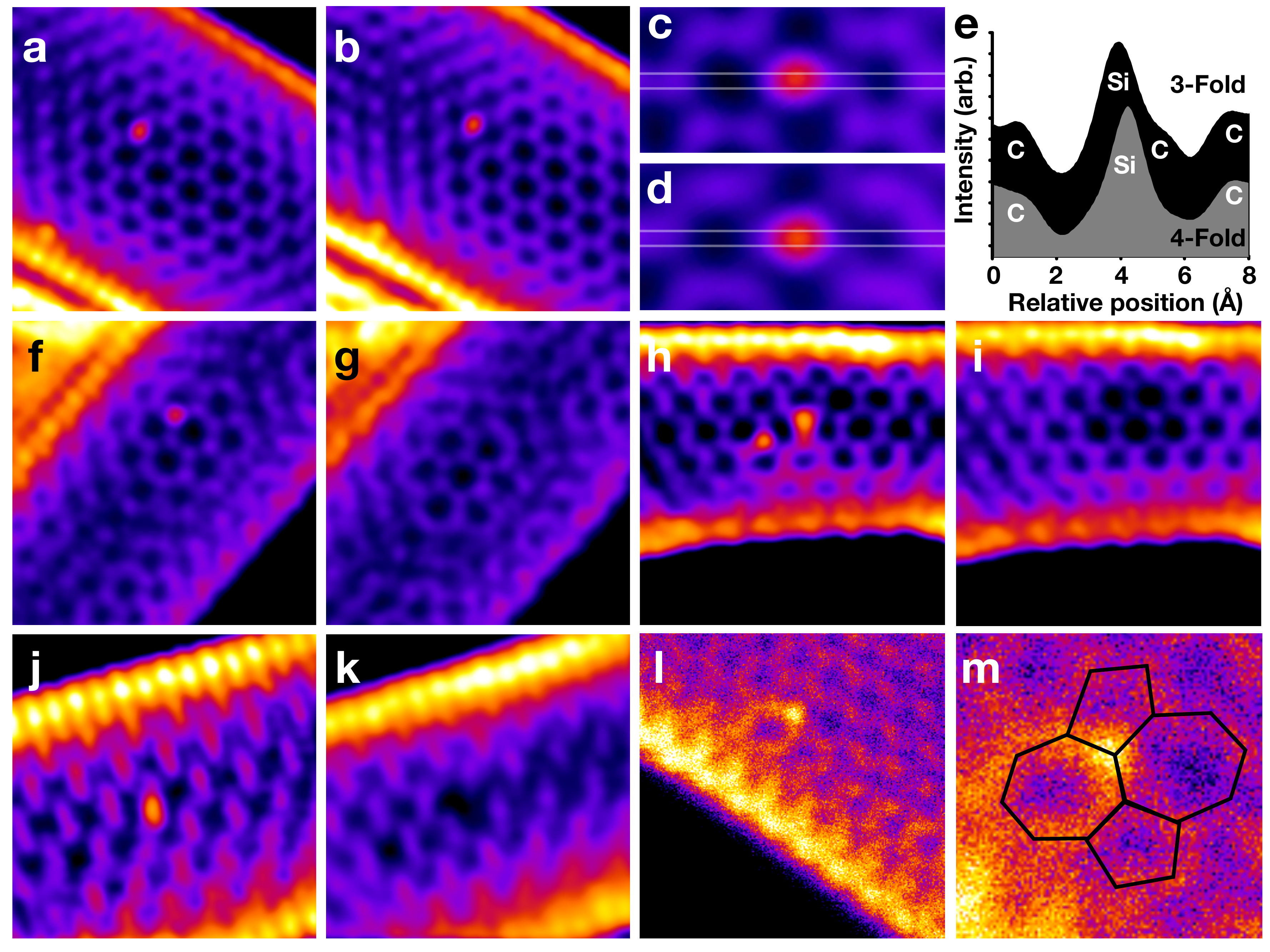}
\caption{Electron-beam induced dynamics. (a-b) Three-to-fourfold conversion of a Si site, with closeups shown in (c-d) and line profiles in (e). A single (f-g) or two (h-i) Si atoms being replaced during manipulation by C. (j-k) An ejection of Si leaving behind a monovacancy. (l-m) Si within a Stone-Wales defect.}
\label{fig:dynamics}
\end{figure}

Jumps were the predominant dynamic in SWCNTs larger than 2~nm in diameter,~\cite{su_competing_2018} but four other kinds of electron-beam induced processes were observed primarily in smaller diameter tubes. The first, ejection of a C neighbour resulting in a three-to-fourfold conversion is shown in \textbf{Figure}~\ref{fig:dynamics}a-b, with close-ups of the local bonding shown in Figure~\ref{fig:dynamics}c-d. Although the contrast is somewhat unclear, line profiles plotted in Figure~\ref{fig:dynamics}e show a change in the relative position of the Si site and the disappearance of the contrast of the C neighbour. Second, we observed the removal of Si atoms during manipulation, which occurs rarely if ever in graphene,~\cite{susi_siliconcarbon_2014} being either replaced by a C atom (Figure~\ref{fig:dynamics}f-g) or leaving behind a monovacancy (Figure~\ref{fig:dynamics}j-k). Interestingly, in rare cases this monovacancy remained stable long enough ($>$4 s) to acquire an image frame. Such vacancies have been reported in graphene,~\cite{Hashimoto_direct_2004,Zhang_scanning_2016,susi_isotope_2016} but are expected to be even more beam-sensitive in SWCNTs~\cite{kotakoski_stability_2012,susi_atomistic_2012} and to our knowledge have not been directly observed. Another peculiar case was observed when trying to move one of two Si atoms bonded within the same hexagon (Figure~\ref{fig:dynamics}h-i): both were replaced by C between the two acquired frames. Finally, we sometimes observed Stone-Wales defects at the Si sites~\cite{tripathi_electron-beam_2018}, such as the one in Figure~\ref{fig:dynamics}l-m.

To evaluate the probability of triggering jumps, we calculated the distribution of electron irradiation doses per event (taking into account that only 26\% of the beam current impinges on the target atom~\cite{tripathi_electron-beam_2018} for our expected probe shape~\cite{kotakoski_imaging_2014}) and estimated the underlying Poisson expectation values by the geometric mean doses.~\cite{susi_siliconcarbon_2014} The dose (geometric mean $\pm$ standard error) required to trigger a jump was (2.1$\pm$0.1)$\times$10$^9$ e$^-$ at 55~kV (N=40) and (6.1$\pm$0.3)$\times$10$^8$ e$^-$ at 60~kV (N=60). These result in cross sections of 0.12~barn at 55~kV and 0.43~barn at 60~kV, slightly higher than the values measured for graphene (0.03~barn at 55~kV and 0.29~barn at 60~kV),~\cite{tripathi_electron-beam_2018} possibly reflecting the slightly curved geometry of even large-diameter SWCNTs. For reasons that will become apparent below, only about half of the 12 separate Si impurities that we attempted to manipulate could be moved.

\begin{figure}[hb!]
\centering
\includegraphics[width=0.55\textwidth,keepaspectratio]{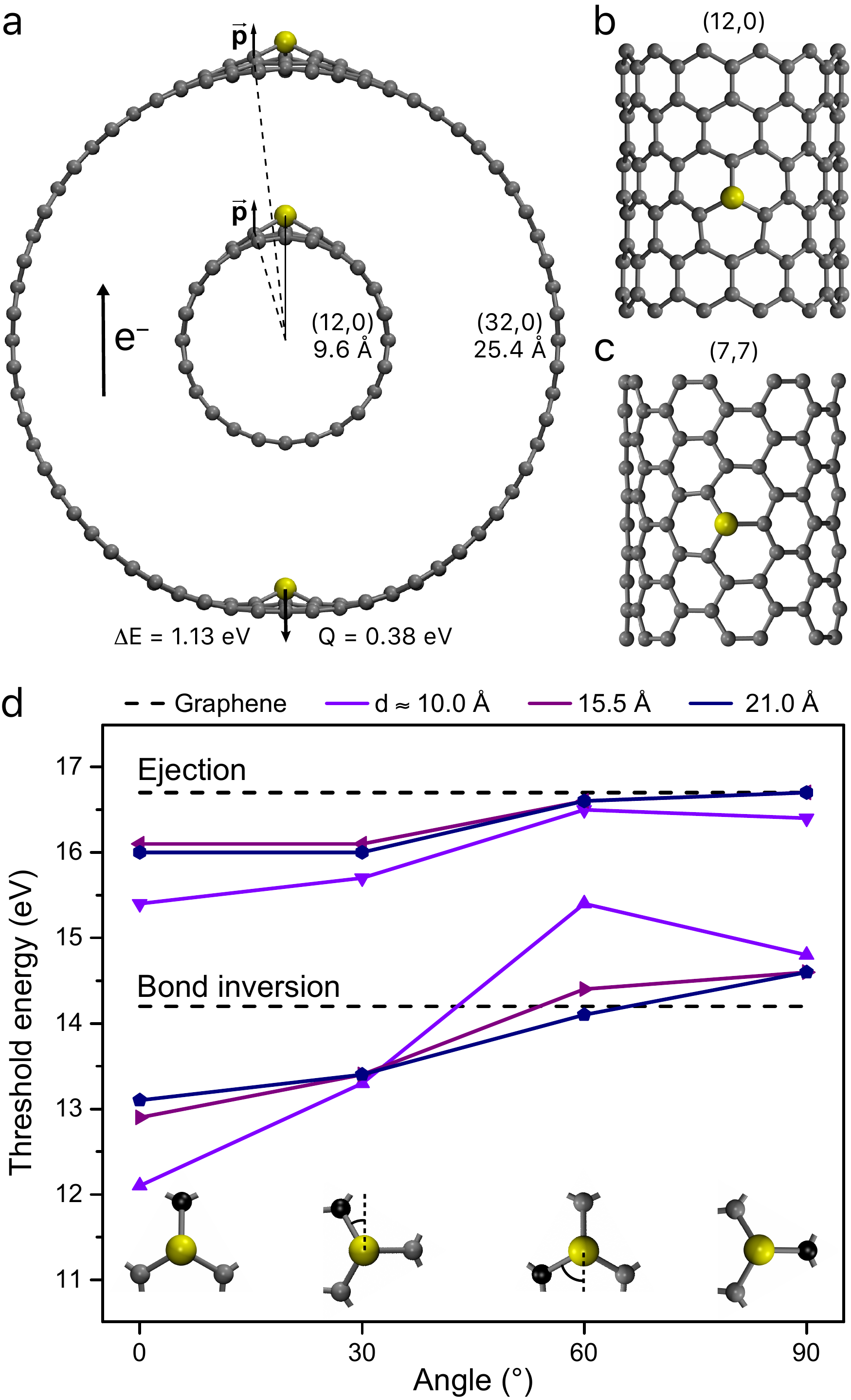}
\caption{The structure and dynamical energetics of Si impurities in SWCNTs. a) Buckling of the Si site in (12,0) and (32,0) nanotubes. The inside configuration shown for the latter is unstable in the former. The dashed lines represent the surface normals at the carbon neighbours, and the directions of the electron beam and the momentum transfer are respectively denoted with e$^-$ and $\mathbf{\vec{p}}$. b) The local configuration of Si-C$_3$ in (12,0) and c) (7,7) nanotubes. d) The energies required for bond inversion and carbon ejection for $\sim$10--21 \AA\ carbon nanotubes for different orientations of the Si-C bonds with respect to the tube axis, as depicted on the inset atomic configurations. Angles 0\textdegree\ and 60\textdegree\ correspond to zigzag tubes, whereas angles 30\textdegree\ and 90\textdegree\ correspond to armchairs. The horizontal dashed lines represent the values for graphene.}
\label{fig:dft}
\end{figure}

To understand the details of the electron-beam induced dynamics, we used DFT/MD simulations,~\cite{susi_isotope_2016} here using the revPBE functional,~\cite{hammer_improved_1999} to study three- and four-fold coordinated Si substitutions in multiple single-walled carbon nanotube models. To account for diameter- and chirality-dependent effects, we considered three armchair (with chiral indices (7,7), (11,11) and (15,15)) and three zigzag (with indices (12,0), (20,0) and (26,0)) tubes, having diameters of approximately 10~\AA, 15.5~\AA\ and 21~\AA. The Si jumps were modeled by running a large number of molecular dynamics simulations in which a single carbon neighbour was provided with an initial momentum ($\vec{p}$) along the direction of the electron beam (denoted with $e^-$ in \textbf{Figure}~\ref{fig:dft}a).

In graphene, a Si substitution buckles about 0.95~\AA\ above the lattice plane due to its larger covalent radius.~\cite{zhou_direct_2012,ramasse_probing_2013,susi_siliconcarbon_2014} In a cylindrical carbon nanotube, the silicon atom could in principle protrude either towards or away from the tube axis, as shown in Figure~\ref{fig:dft}a. Our computations, however, show that the inside configuration is energetically unstable in tubes smaller than $\sim$2~nm in diameter, in which the atom (when purposely placed inside) gets pushed through the lattice during geometry optimization. In over 2~nm tubes, represented by (26,0) and (32,0) chiralities in our simulations, the inside configuration becomes metastable with respective transition barriers of 0.19~eV and 0.38~eV (estimated from single-point energy simulations of eleven images interpolated along the transition path). These transitions, leading to a stable outside configuration, can be thermally activated with respective energy gains of 1.32~eV and 1.13~eV. We can therefore be confident that all Si atoms in our experiments were positioned on the outer side of the nanotube wall. Successful bond inversions in the MD simulations were only achieved for Si atoms positioned on the "backside" of the tube facing away from the electron beam source. The position of forward-facing atoms remains unchanged even when the amount of the energy transferred to a C neighbour is as high as 17~eV. Although the two configurations cannot be distinguished in projected STEM images, this explains why roughly half of the Si atoms failed to move in our experiments.

Additionally, in contrast to graphene where all three C neighbours of the Si are equivalent, in nanotubes this symmetry is broken due to the diameter-dependent curvature. This and the chiral angle affect the length and strain of the C-C (and Si-C) bonds in different directions, as does also the orientation of the Si-C$_3$ site with respect to the nanotube axis. The possible directions of the Si motion correspond to C neighbors at angles (with respect to the tube axis) of 0{\textdegree} and 60{\textdegree} in zigzag and 30{\textdegree} and 90{\textdegree} in armchair tubes, as shown in Fig.~\ref{fig:dft}b-d. Importantly for beam-induced dynamics, since in each case the Si site was aligned with the beam direction, the angle between the (maximum) momentum transferred by an impinging electron and the surface normal varies depending on the tube diameter and the carbon neighbor (Fig.~\ref{fig:dft}a). Further, the alignment of the neighbors of the impacted C atom with respect to the tube circumference influences the restoring forces acting on it. These considerations make different C neighbors nonequivalent and affects the observed dynamics.

The calculated threshold energies for bond inversion and C atom ejection are plotted in Fig.~\ref{fig:dft}d. Interestingly, while the values for 15.5~{\AA} and 21~{\AA} tubes are very similar, apart from the 60{\textdegree} C neighbor, they still considerably differ from the graphene values represented by the horizontal dashed lines. Further, the energies calculated for armchair nanotubes, corresponding to angles 30{\textdegree} and 90{\textdegree}, hardly depend on diameter. This suggests that the angle that the Si-C bond forms with respect to the axis is more important than the nanotube diameter, at least for tubes larger than 10~{\AA}. The most persistent difference from the graphene values is observed for atoms along the tube axis (angles 0{\textdegree} and 30{\textdegree}), which may be due to a weaker restoring force by their C neighbors. The additional divergence of the values for the smallest zigzag (12,0) tube reflects the strong C-C bond strain along its circumference.

Our simulations for even the largest diameter nanotube models show marked anisotropy in the threshold energy values. Moving the Si atom along the axis of either zigzag or armchair tubes requires about 1~eV less energy than in the direction perpendicular to it, which should result in large differences in the observed jump rate at a fixed beam current. However, the experimentally estimated cross sections in the two directions in larger diameter armchair tubes are practically the same within our statistics. One notable disparity with the experiment may be the role of rotation: in the simulations, the impurity site is always perfectly aligned with respect to the beam direction. However, studying this effect systematically is not currently feasible at the required level of theory.

\begin{figure}[t!]
\centering
\includegraphics[width=\textwidth,keepaspectratio]{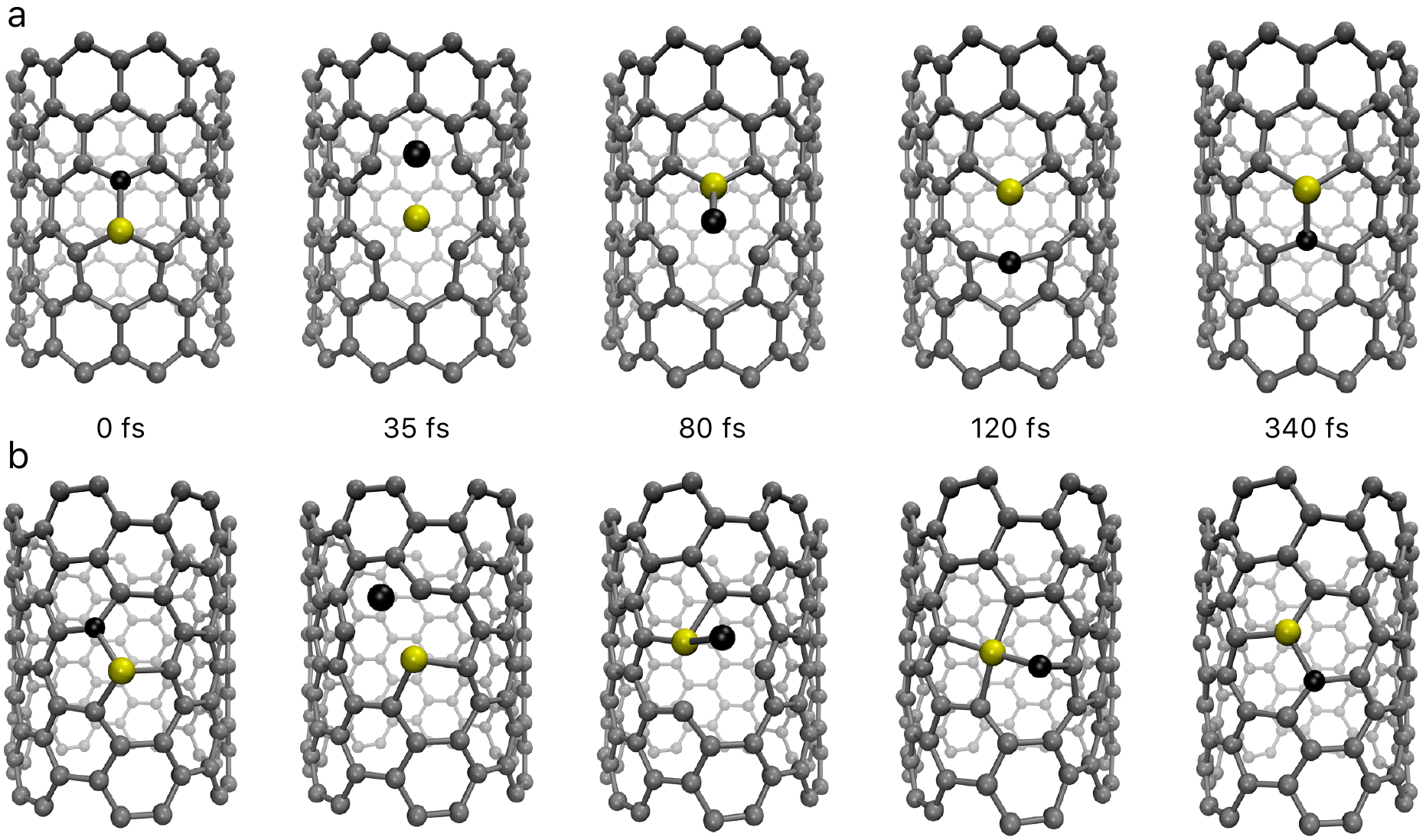}
\caption{Snapshots from DFT/MD bond inversion trajectories. Direct exchange of C and Si taking place along the armchair direction of a (12,0) zigzag nanotube in a) and the zigzag direction of a (7,7) armchair tube in b) after an energy transfer of 14 eV to the C atom shown in black.}
\label{fig:dft2}
\end{figure}

While the observed bond inversion process in SWCNTs is fundamentally similar to graphene, we did notice some minor differences. \textbf{Figure}~\ref{fig:dft2}a shows the Si-C bond inversion along the (12,0) zigzag tube axis. In such cases, the C neighbours of the impacted atom are symmetric with respect to the inverted Si-C bond and the tube circumference, and thus the trajectory closely resembles that in graphene.~\cite{susi_towards_2017} This is however not the case for the bond inversion at sites located at 30{\textdegree} and 60{\textdegree} angles, where the symmetry of the neighbours breaks down, resulting in a curved trajectory in which the C atom does not directly cross over the Si but appears to traverse "around" it, resembling the mechanism of the Stone-Wales transformation in graphene.~\cite{kotakoski_stone-wales-type_2011} An example of a 30{\textdegree} case is presented in {Figure}~\ref{fig:dft2}b, depicting a bond inversion in a (7,7) zigzag tube with an initial C atom kinetic energy of 14 eV.

The curved trajectories, in turn, may result in an incomplete reconstruction and appearance of defective structures. For example, at energies lower than those required for bond inversion, the C neighbour sometimes ends up as an adatom bound either on a Si-C or C-C bridge near the site of the electron impact, resulting in a fourfold-coordinated Si. In our simulations, the thresholds for such structures to emerge in zigzag nanotubes were 13.4--13.7 eV and in armchair tubes 11.0--11.9 eV, the process in smaller tubes requiring less energy. The recombination of the four-fold coordinated Si with the C adatom results in a significant gain of energy similar to graphene,~\cite{susi_siliconcarbon_2014} and therefore we assume that these structures are metastable and will eventually reconstruct into Si-C$_3$. Due to computational constraints, however, it was not possible to simulate the dynamics long enough to observe this process directly, except for several cases in the (15,15) tube where the recombination did indeed occur during the MD simulations. We also note that for curved trajectories, we observed several cases of Stone-Wales transformations~\cite{kotakoski_stone-wales-type_2011} at the Si site~\cite{tripathi_electron-beam_2018} within the energy ranges that would otherwise correspond to bond inversions.

In summary, we have demonstrated via atomically resolved scanning transmission electron microscopy experiments that silicon impurities in single-walled carbon nanotubes can be manipulated with focused electron irradiation. The curved geometry of the tubes affects the threshold energies as predicted by density functional theory molecular dynamics simulations. The orientation of the silicon site with respect to the beam direction plays an important role: only impurities along the axis can be easily visualized, and only those incorporated into the wall facing away from the beam can be moved. Further, the computations predict an asymmetry in the armchair and zigzag directions for smaller diameter tubes, although in our experiments with large diameter tubes we did not observe this. Our results expand the range of strongly bound materials where impurities can be manipulated with a focused electron beam beyond two dimensions,~\cite{jesse_direct_2018,hudak_directed_2018} further underscoring the potential of scanning transmission electron microscopy as a tool for atomically precise manipulation. Controllable modification of the quasi-1D electronic structure of single-walled carbon nanotubes may enable devices such as tunable single-molecule Fabry-P\'erot interferometers and other electron wave resonators.~\cite{liang_fabry_2001}

\subsubsection*{Experimental Section}
\textit{Sample preparation} The single-walled carbon nanotubes were synthesized in a vertical floating catalyst reactor using ethanol (C$_2$H$_5$OH) as the primary carbon source and hydrogen (H$_2$) as a reaction mediator.~\cite{ding_highly_2017,ding_high-performance_2018} Ferrocene and thiophene were used as a catalyst source and growth promoter, respectively. The synthesis conditions were chosen to favour tubes with high helicities and large diameters, feeding 300~cm$^3$/min of H$_2$ and 300~cm$^3$/min of nitrogen carrying C$_2$H$_5$OH at the rate of 4~$\mu$l/min. The material was deposited by placing a perforated silicon nitride membrane acquired from Ted Pella Inc. on a membrane filter, through which the reactor exhaust was passed for $60-120$~s, accumulating nanotube networks suitable for high-resolution electron microscopy.

The silicon (Si) substitution was carried out in a custom-made plasma chamber connected to the electron microscope through an ultra-high vacuum transfer system.~\cite{inani_silicon_2019} Ar$^+$ ions, formed in a microwave plasma cavity at the pressure of $\sim 5\times 10^{-6}$ mbar, were accelerated to a kinetic energy of $\sim 50$~eV to create intermittent vacancies in the tube walls.~\cite{inani_silicon_2019} The total radiant exposure was $\sim$1~ion~nm$^{-2}$. Concurrently, the samples were irradiated with a high-power laser similar to the one we previously used for cleaning 2D materials,~\cite{tripathi_cleaning_2017} which in this case thermally mobilized Si impurity atoms to fill the created vacancies. After plasma irradiation, the samples were transferred directly to the electron microscope in ultra-high vacuum without exposing them to the ambient atmosphere.

\textit{Scanning transmission electron microscopy} 
All experiments were conducted using the aberration-corrected Nion UltraSTEM100 scanning transmission electron microscope in Vienna, operated at two electron energies, 55 and 60~keV, in ultra-high vacuum (10$^{-10}$ mbar). The typical beam current of the instrument is close to 35~pA. The beam convergence semi-angle was 30 mrad and all images were acquired with the medium angle annular dark field (MAADF) detector with an semi-angular range of 60--200 mrad. Electron energy loss spectroscopy was used to identify the Si impurities (for details, see Ref.~\citenum{inani_silicon_2019}). To remove the influence of probe tails, some images were processed using a double Gaussian filtering procedure~\cite{krivanek_atom-by-atom_2010} and colored with the ImageJ lookup table ”fire” to enhance contrast.

\textit{Density functional theory} 
All simulations were performed using density functional theory (DFT) as implemented in the GPAW package.~\cite{enkovaara_electronic_2010} We used the revised PBE~\cite{hammer_improved_1999} exchange-correlation functional, a localized (\textit{dzp}) basis set,~\cite{larsen_localized_2009} a grid spacing of 0.2~\AA, and three \textbf{k}-points in the periodic axial direction. The length of the armchair and zigzag SWCNTs was 12.35 and 12.83~\AA. The supercell size in the directions perpendicular to the tube axes has been set to 38~\AA, which gives more than 10~\AA\ separation between the periodic replicas even for the largest considered (32,0) SWCNT with a diameter of 25.4~\AA.

The displacement threshold energies were calculated using Velocity-Verlet molecular dynamics~\cite{larsen_atomic_2017} with a time step of 0.5 fs and varying the amount of the energy transferred to a C atom at 0.1 eV intervals (a detailed description of the methodology can be found in Ref.~\citenum{susi_isotope_2016}). We used a longer time step here (0.5 fs instead of 0.1 fs) to facilitate the large number of required simulations as well as the nanotube models with up to 312 atoms. Our threshold energy for pristine graphene is about 3\% greater than reported earlier,~\cite{susi_isotope_2016} and test simulations for the (7,7) tube with respect to the time step, tube length and \textbf{k}-points show that the presented values of the threshold energies are converged to within 0.3~eV.

%\subsubsection*{Supporting Information}
%Supporting Information is available from the Wiley %Online Library or from the authors.

\subsubsection*{Acknowledgements}
M.T. and T.S. acknowledge the Austrian Science Fund (FWF) project P 28322-N36 for funding, and T.S. and A.M. also the European Research Council (ERC) Grant No. 756277-ATMEN. E.I.K, E.D. and A.H. acknowledge the support from the Academy of Finland \textit{via} projects 286546-DEMEC and 292600-SUPER, from TEKES Finland \textit{via} projects 3303/31/2015 (CNT-PV) and 1882/31/2016 (FEDOC), and the Aalto Energy Efficiency (AEF) Research Program through the MOPPI project. A.M. and T.S. acknowledge the Vienna Scientific Cluster for computer time. J.K., H.I. and K.M. were supported by the FWF project I3181 and the Wiener Wissenschafts\mbox{-,} Forschungs- und Technologiefonds (WWTF) project MA14-009, and K.M. further by the Finnish Cultural Foundation through a grant from the Finnish Postdoc Pool and J.K. through FWF project P31605.

\subsubsection*{References}
\renewcommand{\section}[2]{}%

\end{document}